\documentclass[epj,referee]{svjour}
\usepackage{amsmath,amssymb,graphicx}
\usepackage{color,ulem,latexsym}

\begin{document}
\title{Birth and Death in a Continuous Opinion Dynamics Model. The consensus
  case} 
\author{Timoteo Carletti\inst{1}, Duccio Fanelli\inst{2,4}, Alessio
  Guarino\inst{3}, Franco Bagnoli\inst{4} \and Andrea Guazzini\inst{4}
}                     
%
%
\institute{1. D\'epartement de Math\'ematique, Facult\'es Universitaires Notre
  Dame de la 
  Paix, 8 rempart de la vierge B5000 Namur, Belgium \email{timoteo.carletti@fundp.ac.be}\\
2. School of Physics and Astronomy, University of Manchester, M13 9PL,
    Manchester, United Kingdom \email{duccio.fanelli@manchester.ac.uk}\\
3. Universit\'e de la Polyn\'esie Francaise, BP 6570 Faa'a, 98702, French
  Polynesia \email{alessio.guarino@upf.pf}\\
4. Dipartimento di Energetica and CSDC, Universit\`a di
  Firenze, and INFN, via S. Marta, 3, 50139 Firenze, Italy
  \email{franco.bagnoli@unifi.it} \email{andrea.guazzini@unifi.it}}
\date{Received: \today}
%
\abstract{We here discuss the process of opinion formation in an open
community where agents are made to interact and
consequently update 
their beliefs. New actors (birth) are assumed to replace individuals
that abandon the community (deaths). This dynamics
is  
simulated in the framework of a simplified model that accounts for mutual
affinity between agents. A rich phenomenology is presented and
discussed  
with reference to the original (closed group) setting. Numerical findings are
supported by analytical calculations. 
} 
\authorrunning{Timoteo Carletti et al.}
\maketitle

\section{Introduction}

Opinion dynamics modeling represents a challenging field where ideas from statistical physics and non-linear science 
can be possibly applied to understand the emergence of collective behaviors,
like consensus or polarization in social groups. Several models 
have been proposed in the past to reproduce the key elements that
supposedly drive the process of opinion making~\cite{opinion,baronchelli}. 
The problem of providing an adequate experimental
 backup to such developments is indeed an open one, and more work is
 certainly needed to eventually assess the  interpretative ability of the
 proposed mathematical 
formulations, following e.g. the guidelines of
 ~\cite{galamepjb,galamejsp}. Models based on interacting agents
display 
however a rich and intriguing dynamics which deserves to be fully unraveled.

Opinion dynamics models can be classified in two large groups. On the one hand, opinions are represented as discrete (spin--like)
variables where the system behaves similarly to spin glasses
models~\cite{Sznajd}. On the other, each individual bears a continuous
opinion   
which span a pre--assigned range~\cite{Deffuant}. More recently, a new
framework for a discrete but unbounded number of opinions was proposed
in \cite{baronchelli} and shown to nicely complement the picture. In
all the above
approaches, a closed system is generally assumed, meaning that the same pool of actors is made to interact during the 
evolution. This can be interpreted by 
  assuming that the inherent dynamical timescales (e.g. opinion convergence
time) are 
much faster than those associated to the processes (e.g. migration,
birth/death) responsible for a modification of the group
composition, 
these latter effects having being therefore so far neglected. Such an implicit assumption is certainly correct when the debate is bound to a small community of individuals, thus making it possible 
to eventually achieve a rapid convergence towards the final configuration. Conversely, it might prove inaccurate when applied to a large ensemble of interacting 
agents, as the process becomes considerably slower and external perturbations need to be accounted for. Given the above, it is therefore of interest 
to elucidate the open system setting, where the population is periodically renewed. 

To this end, we refer to the model presented in~\cite{Bagnoli_prl}, where the
role of affinity among individuals is introduced as an additional ingredient.  
This novel quantity measures the degree of inter--personal intimacy and
sharing, an effect of paramount importance in real social system~\cite{Nowak}
. Indeed, the outcome of an  
hypothetic binary interaction relies on the difference of opinions, previously
postulated, but also on the quality of the mutual relationships.  
The affinity is dynamically coupled to the opinion, and, in this respect, it
introduces a memory bias into the system: affinity between agents increases
when their opinion tends to converge.  

The aforementioned model is here modified to accommodate for a
death/birth like process. In this formulation, $M$ agents
are randomly eliminated from the system, every $T$ time steps. When an
agent exits from the community (virtually, dies), he is immediately replaced
by a new element,   
whose opinion and affinity with respect to the group are randomly assigned. As
we shall see, the perturbation here prescribed alters dramatically the
behavior of the system, with reference to  
the ideal close--system configuration. To understand such modifications via
combined numerical and analytical tools, constitutes the object of   
the investigations here reported.

We shall mainly explore the parameters setting that would lead to an asymptotic 
consensus state (all agents eventually bearing the shared opinion $0.5$), in absence of the external 
perturbation which is here object of investigations. 
The underlying model~\cite{Bagnoli_prl} displays however a richer phenomenology, exhibiting in particular stable polarized  
states in the late time evolution and long-lived metastable regimes. Though it would be extremely interesting to extend 
the present analysis and hence cover those additional scenarios, we chose to only briefly touch upon this issue when commenting the
details of the transition between single and fragmented phases.

The paper is structured as follows. We first introduce the model, then we
present the obtained analytical and numerical results and, finally, we sum up
and draw our conclusions.


\section{The model}
\label{sect:model}

In the following, we will review the model previously introduced in ~\cite{Bagnoli_prl} and present the additional features
that are here under inspection. The interested reader can thus
refer to the original paper \cite{Bagnoli_prl}  for an additional account on the model characteristics.

Consider a population of $N$ agents and assume that at time $t$ they bear a  scalar opinion $O_i^{t} \in [0,1]$. We also introduce the $N \times N$ time 
dependent matrix ${\bf \alpha}^{t}$, whose elements $\alpha_{ij}^{t}$ belong to the interval $[0,1]$. The quantities $\alpha_{ij}^{t}$  specify the affinity 
of individual $i$ vs. $j$, at time $t$: Larger values of  $\alpha_{ij}^{t}$ are associated to more trustable relationships. 
    
Both the affinity matrix and the agents opinions are
randomly initialized 
time $t=0$.  At each time step $t$, two agents, say $i$ and $j$, are chosen
according to the following extraction rule: first the agent $i$ is randomly identified, with a
uniform probability. Then, the agent $j$ which is closer to $i$ in term of the social metric 
$D_{ij}^\eta$  is selected  for interaction. 
The quantity $D_{ij}^\eta$ results from the linear superposition of the so--called social distance,  $d_{ij}$,  
and a stochastic contribution $\eta_j$, namely:
\begin{eqnarray}
 D_{ij}^\eta = d_{ij}^t + \eta_j(0,\sigma)\, .
\label{social_metrics}
\end{eqnarray}

Here $\eta_j(0,\sigma)$ represents a 
normally distributed noise, 
with mean zero and variance $\sigma$, the latter being named social temperature.  
The social distance is instead defined as:
\begin{eqnarray}
d_{ij}^t &=& \Delta O_{ij}^{t} (1-\alpha_{ij}^{t}) \qquad j=1,...,N \qquad j
\ne i\, ,
\label{social_distance}
\end{eqnarray}
with $\Delta O_{ij}^{t}= O_i^t-O_j^t$.

The smaller the value of $d_{ij}^t$ the closer the agent $j$ to $i$,
both in term of affinity and opinion. The additive noise $\eta_j(0,\sigma)$  
acts therefore on a fictitious 1D manifold, 
which is introduced to define the pseudo--particle (agent) interaction on the
basis of a nearest neighbors  
selection mechanism and, in this respect, set the degree of mixing in the
community.

When the two agents $i$ and $j$ are extracted on the basis of the recipe
prescribed above, they interact and update their characteristics according to
the following scheme~\footnote{The evolution of the quantities $O_j(t)$ 
  and $\alpha_{ji}(t)$ 
is straightforwardly obtained by switching the labels $i$ and $j$ in
the equations.}:  
\begin{equation}
\label{opinion}
  \begin{cases}
O_i^{t+1} &= O_i^{t}- \frac{1}{2} \Delta O_{ij}^{t}
\Gamma_1\left(\alpha^t_{ij}\right) \\ 
\alpha_{ij}^{t+1} &= \alpha_{ij}^{t} + \alpha_{ij}^{t}
      [1-\alpha_{ij}^{t}] \Gamma_2 \left(\Delta O_{ij}\right) \, ,
  \end{cases}
\end{equation}
where the functions $\Gamma_1$ and $\Gamma_2$ respectively read:
\begin{equation}
  \label{eq:Gamma1}
  \Gamma_1 \left(\alpha^t_{ij}\right)= \Theta \left(
  \alpha_{ij}^{t}-\alpha_c\right)
\end{equation}
and
\begin{equation}
  \label{eq:Gamma2}
  \Gamma_2 \left(\Delta O_{ij}\right)= \Theta\left( \Delta
O_c-|\Delta O_{ij}^{t}|\right) - \Theta\left(|\Delta O_{ij}^{t}| - \Delta
O_c\right)
\end{equation}
and the symbol $\Theta(\cdot)$ stands for the Heaviside
step--function~\footnote{In   
\cite{Bagnoli_prl}, the switchers $\Gamma_1$ and $\Gamma_2$ are smooth
functions constructed from  
the hyperbolic tangent. We shall here limit the discussion to considering the
Heaviside approximation, which is  
recovered by formally sending $\beta_{1,2}$ to infinity in Eqs. (3) and (4)
of~\cite{Bagnoli_prl}.}. More specifically,    
$\Gamma_{1}$ is $0$ or $1$ while $\Gamma_{2}$ is $-1$ or 
$1$, depending on the value of their respective arguments.  In the above
expressions $\alpha_c$ and $\Delta
O_c$ are constant 
  parameters. Notice that, for $\alpha_c
  \rightarrow 0$, the opinion is formally decoupled from affinity
  in~\eqref{opinion} being $\Gamma_1 = 1$ irrespectively of the actual value
  of $\alpha_{ij}^t$, and the former evolves following the Deffuant et 
  al. scheme~\cite{Deffuant} with convergence rate $\mu=0.5$ and interaction
  threshold 
    $d=1$ (confidence bound). The latter scheme pioneered the broad 
class of models inspired to the so--called bounded confidence
hypothesis, an assumption which, though revisited, also enters the  
self-consistent scenario of Eqs.\eqref{opinion}.  
 
\begin{figure} 
\centering
\includegraphics[width=7cm]{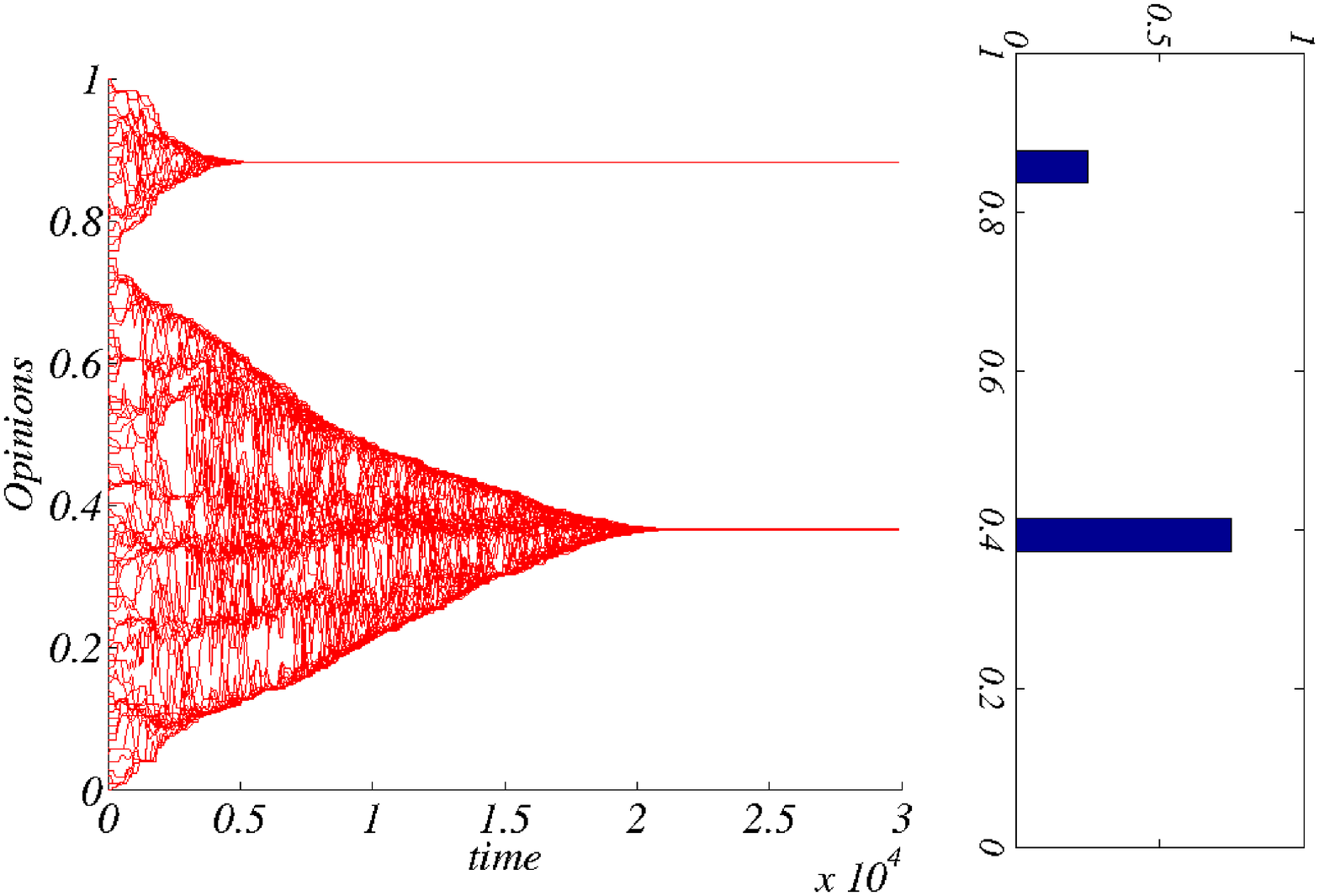}
\includegraphics[width=7cm]{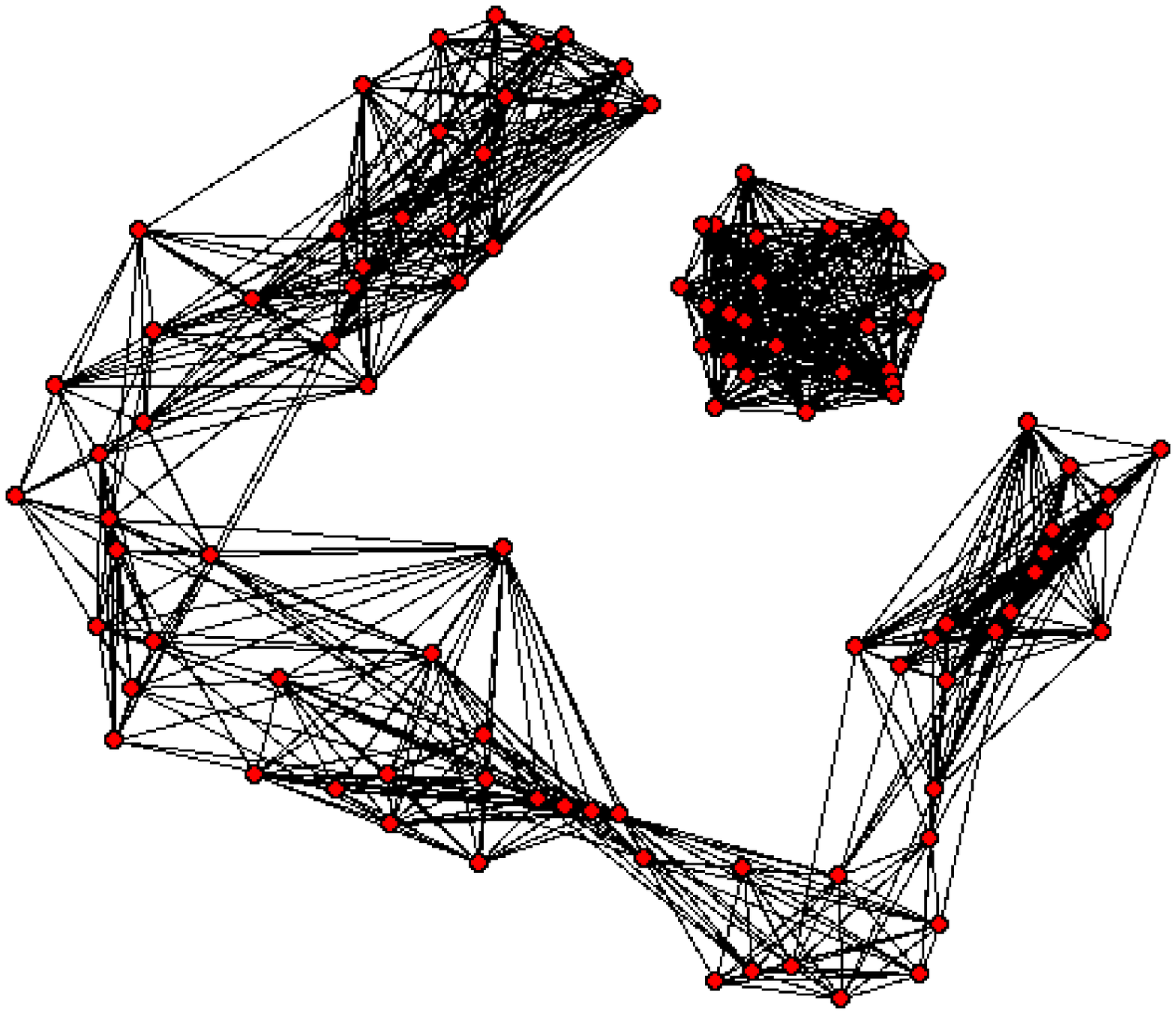}
\caption{Upper panel: Typical evolution of the opinion
  versus time, 
i.e. number of iterations and the asymptotic distribution of agents' opinion
in a closed community. Bottom panel: Final
affinity matrix, represented by the underlying network: node $i$ is linked to
node $j$ if $\alpha_{ij} > \bar{\alpha}=0.8$. Here 
$\sigma= 4 \cdot 10^{-4}$, $\Delta O_c = 0.5$, $\alpha_c = 0.5$ and
$\rho=0$, i.e. no agent can leave the group. Initial opinion are
(random) uniformly distributed within the interval $[0, 1]$, while
$\alpha_{ij}^0$ 
is initialized with uniform (random) values between $0$ an $0.5$.}
\label{fig_qualitativa}
\end{figure}

In ~\cite{Bagnoli_prl}, a preliminary analysis of the qualitative behavior of
the model as a function of the involved parameters is reported. Asymptotic 
clusters of opinion are formed, each agglomeration being different in size and
centered  
around distinct opinion values. Individuals sharing the same believes are also 
characterized by a large affinity scores, as it is exemplified in 
Fig.~\ref{fig_qualitativa}.  

More quantitatively, the system is shown to undergo a continuous phase transition: above a critical value of the control parameter
$(\sigma \alpha_c)^{-1/2}$ it fragments into several opinion clusters, otherwise convergence to a single group is numerically 
shown to occur \footnote{Strictly speaking, it should be noted that the fragmented state is metastable, if the mean separation between the adjacent peaks 
is smaller than the interaction distance $\Delta O_c$. There always exists in fact a finite, though small, probability of selecting two individuals 
which belong to different agglomerations. When the above condition applies, i.e. when the agents' opinions are closer than the threshold amount  $\Delta O_c$, such rare 
encounters produce a gradual increase of the mutual affinity scores, a tendency which asymptotically drives a merging of the segregated 
clusters, as ruled by Eqs. (\ref{opinion}). This final convergence is eventually achieved on extremely long time scales, diverging with the number of agents. 
Socially relevant dynamics are hence bound to the metastable phases, which are being investigated in Fig.~\ref{fig1}.}.  
We shall here simply notice that a significant degree of mixing (large social
temperature $\sigma$) brings the system towards the single--cluster 
final configuration.   

Starting from this setting, we introduce the birth/death process, which in turn amounts to place the system in contact with an external reservoir. The perturbation
here hypothesized is periodic and leaves the total number of agent unchanged. Every $T$ time steps (i.e. encounter events) $M$ agents, randomly selected, are forced to
abandon the system (death). Every removed individual is instantaneously replaced by a new element, whose initial opinion and affinity are randomly fished, with
uniform probability, from the respective intervals $[0,1]$ and $[0,
\alpha_{max}]$. Further, we introduce $\rho=\frac{M}{T}$ to characterize the
{\it departure density}, a crucial
quantity that will play the role of the control parameter in our
subsequent developments. As a final remark, it should be emphasized that no aging mechanisms 
are introduced: agents are mature enough to experience peer to peer encounters from the time they enter the system.

\section{Results}
\label{sect:result}

\begin{figure}[htbp]
\centering
\includegraphics[width=9cm]{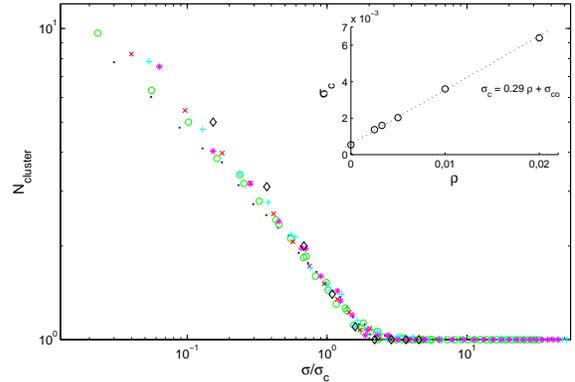}
\caption{Main plot: Average number of clusters as function of the rescaled
  quantity $\frac{\sigma}{\sigma_c}$ for different values of the density   
$\rho$. Simulations have been performed with parameter values: $\rho=0$
($\Diamond$), $\rho=0.0025$ ($\ast $), $\rho=0.0033$ ($+$), $\rho=0.005$
($\times $), $\rho=0.01$ ($\circ $) and $\rho=0.05$ ($\blacklozenge$). In
all simulations, here and after reported, unless otherwise specified,   
$\alpha_c = \Delta O_c = 0 .5$; $O_i^0$ and $\alpha_{ij}^0$ are random
variables uniformly distributed in the intervals $[0,1]$ and  
$[0,\alpha_{max}]$ -- being $\alpha_{max}=0.5$ -- respectively. Inset :
$\sigma_c$ as a function of $\rho$. The open circles, $\circ$, represent the values
calculated from the transitions shown in the main plot. The dotted lines
represent the best linear fit : $\sigma_c=0.3 \rho + \sigma_{co}$, with
$\sigma_{co}=5.5 \cdot 10^{-4}$. Notice that $\sigma_c=\sigma_{co}$ is
eventually recovered in the closed-system setting, which  
in turn corresponds to $\rho=0$.}  
\label{fig1}
\end{figure}

Numerical simulations are performed for a system of $N=100$ individuals and 
its evolution monitored  \footnote{
The chosen value of $N$ could be in principle considered too small to allow us extracting sound
statistical information from the model at hand. As we shall however discuss, already at such 
relatively small value of $N$, one observes a satisfying matching between numerics and statistical based 
predictions. No substantial differences are detected when simulating a larger system, this observation motivating our  
choice to  stick to the $N=100$ case study. Notice also that potentially interesting applications in social sciences would often deal with 
a finite, possibly limited, number of agents, as for the case being addressed at present.}.  
Qualitatively, the system shows the typical critical
behavior  as observed  
in the original formulation \cite{Bagnoli_prl}. However, peculiar distinctions are found, some of those being addressed in the
following discussion. 
First, an apparently smooth transition is also observed within this novel
formulation, which divides the mono-- and multi--clustered
phases. 
Interestingly,  
the transition point is now sensitive to the departure density $\rho$. To
further elucidate this point, we draw in Fig.~\ref{fig1} the average number
of  
observed clusters versus a rescaled temperature. A clear transition towards
an ordered (single-clustered) phase is observed,  
as the temperature increases. The parameter $\sigma_c$ in Fig.~\ref{fig1} plays the role of
an effective temperature, and it is numerically 
adjusted to make distinct curves collapse onto the same profile, which hence
applies to all values of $\rho$. The inset of Fig.~\ref{fig1}
shows that there is a 
linear correlation between $\sigma_c$ and $\rho$. The larger the departure density $\rho$, the larger the effective temperature $\sigma_c$. In other words, 
when $\rho$ is made to increase (i.e. the system is experiencing the effect of a more pronounced external perturbation), one needs to augment the degree of mixing,
here controlled by the social temperature $\sigma$, if a convergence to the final mono--cluster is sought.  
The death/birth process is in fact acting
against the thermal contribution, which  brings into contact otherwise  
socially distant individuals. While this latter effect enhances the chances of
convergence, the former favors the opposite tendency to spread.

\begin{figure*} 
\centering
\includegraphics[width=15cm]{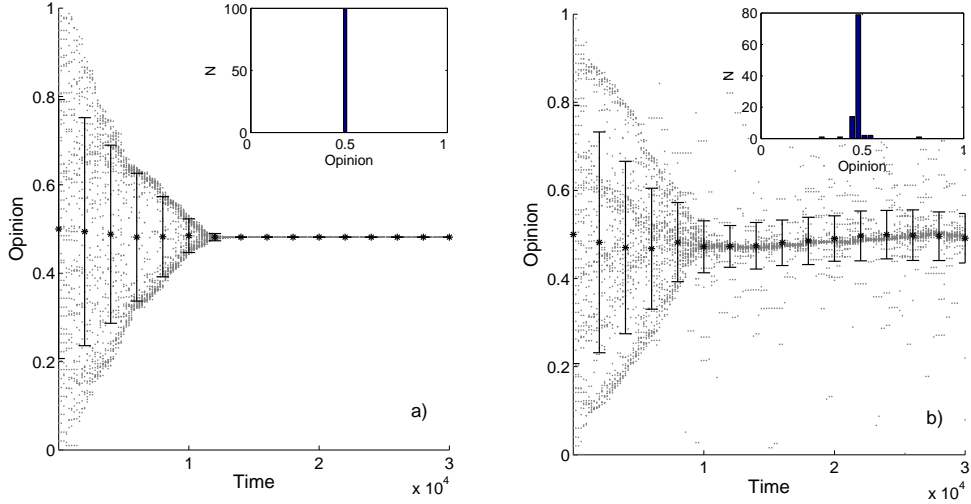} 
\caption{Opinion evolution versus
  time, the latter being quantified through the number of iterations. The
  black stars represent the mean opinion and the error bar the standard
  deviation of the opinion distribution. In the insets, the histogram of
  asymptotic 
  distribution of agents' opinion.  
In panel a) the system is closed (i.e. $\rho=0$), in b) $\rho=5 \cdot
10^{-3}$.}  
\label{fig2}
\end{figure*}

To further elucidate the role of the external perturbation, we shall refer to
the dynamical regime where the agents converge to a  single  
cluster. When $\rho$ is set to zero, the final shared opinion is 0.5 to which
all agents eventually agree, see Fig.~\ref{fig2}a. In other words, the final 
distribution is a Dirac delta, with the peak positioned in
$O=0.5$. Conversely, for positive, but small, values of $\rho$, 
 the final distribution of opinions presents
a clear spreading around  
the most probable value, still found to be $0.5$. This scenario is clearly
depicted in Fig.~\ref{fig2}b. For larger $\rho$, when the birth-death perturbation becomes more frequent, the opinion profile cannot
relax away from the initial distribution, the agents believes being uniformly scattered over the allowed interval, i.e. $[0,1]$. 

\begin{figure}[htbp]
\centering
\includegraphics[width=9cm]{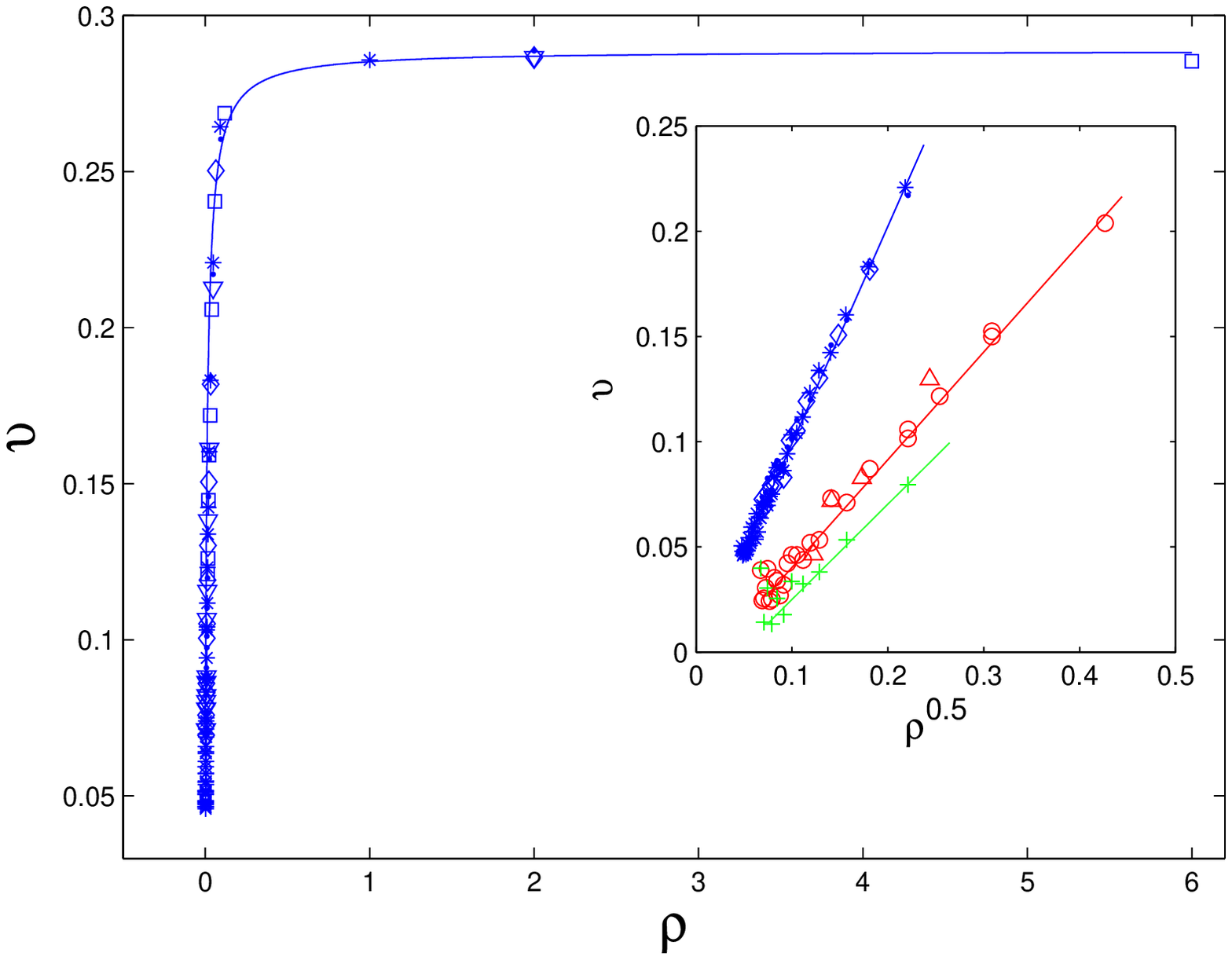}
\caption{Main: The
  standard deviation of the final mono--cluster $\upsilon$ as 
  a function of the departure density $\rho$. Each point results from
  averaging out $100$  
independent runs, with $N=100$. Symbols refer to numerical
simulation performed with 
different values of $\alpha_c$, $M$ and $\sigma$ : ($\ast$) $\alpha_c=0.5, M=1,
\sigma=0.07$, ($\Diamond$) $\alpha_c=0.5$, $M=2$, $\sigma=0.07$, ($\Box$)
$\alpha_c=0.5$, $M=6$, $\sigma=0.07$, ($\bullet$) $\alpha_c=0.5$, $M=2$,
$\sigma=0.25$, ($\bigtriangledown$) $\alpha_c=0.5$, $M=2$, $\sigma=1$, ($\circ$)
 $\alpha_c=0.3$, $M=2$, $\sigma=0.28$, ($\bigtriangleup$) 
$\alpha_c=0.3$, $M=6$, $\sigma=0.28$ and ($\times$) $\alpha_c=0$, $M=2$,
$\sigma=0.28$). The solids line refers to the theoretical
prediction~\eqref{standard}: The free parameter $T_c$ is numerically fitted
and results in $T_c= 8030$ for $\alpha_c=0.5$, $T_c= 1886$ for $\alpha_c=0.3$
and $T_c= 1470$ for $\alpha_c=0$. This values are in good agreement with the
simulation results~\cite{tempoconv}, which confirms the validity of the
proposed analytical scheme.  Inset: $\upsilon$ versus $\sqrt{\rho}$, for $\rho
\in [0,0.5]$. This zoomed view confirms   
the correctness of the scaling dependence derived in~\eqref{small_rho}.} 
\label{fig3}
\end{figure}

The associated standard deviation $\upsilon$ is deduced, from a series of simulations, and shown to depend on the selected value of $\rho$. The result of the analysis 
is reported in Fig.~\ref{fig3}, where the calculated value of $\upsilon$ (symbols) is plotted versus the departure density amount
$\rho$. 

For small values of the control parameter $\rho$, the standard deviation $\upsilon$ of the cluster scales proportionally to $\sqrt{\rho}$, Fig.~\ref{fig2}. 
Numerics indicate that the proportionality coefficient gets smaller, as $\alpha_c$ grows. In the opposite limit, namely for large values of the density 
$\rho$, the standard deviation $\upsilon$ rapidly saturates to a asymptotic value, $\upsilon_c$. The latter is universal, meaning that it neither scales with 
$\alpha_c$, nor it does with the social temperature $\sigma$. Our best
numerical estimates returns, $\upsilon_c =0.28 \simeq 1/\sqrt{12}$ which, as
expected, corresponds to the standard deviation of the uniform distribution in
the interval of  
length 1.  

The solid lines in Fig.~\ref{fig3} represent the function :
 \begin{eqnarray}
\upsilon^2  = \frac{M}{12 N [1- \frac{N-M}{N} (\frac{T_c-T}{T_c})^2 ]
  }\, ,
 \label{standard}
\end{eqnarray}
which straightforwardly follows from an analytical argument,
developed 
hereafter. In the above expression $T_c$, stands for an effective estimate for   
time of convergence of the opinion cluster, and is deduced
  via numerical fit (see caption of Fig.~\ref{fig3} and~\cite{tempoconv} for
  further details).    
In~\cite{propaganda}, working
within the Deffuant's scheme~\cite{Deffuant}, i.e. closed
  community case without affinity, the time needed 
to form a coherent assembly from a sequence of binary encounters  
was shown to diverge with the population size
$N$, with a super-linear scaling. Moreover,  it
was also proven that the affinity slows down the convergence rate, a
fact that can be successfully captured by accounting for an additional dependence of $T_c$ over 
$\alpha_c$: The larger $\alpha_c$ the longer the convergence time, as reported in~\cite{Bagnoli_prl}. A comprehensive discussion on the
analytical derivation of $T_c(\alpha_c)$ falls outside the scope of the present discussion and will be 
presented in a forthcoming contribution~\cite{carletti}.

Before
turning to discuss the analytical derivation of
Eq.~\eqref{standard}, we wish to test its predictive adequacy with
reference to the two limiting cases outlined above. Indeed, for $\rho<<1$ and
$T<<T_c$,  Eq.~\eqref{standard} can be cast in the approximated form:
 \begin{eqnarray}
\upsilon = \sqrt{\frac{T_c (\alpha_c)}{24 N}} \sqrt{\frac{M}{T}} =
\gamma_t \sqrt{\rho}\, ,  
 \label{small_rho}
\end{eqnarray}
which presents the same dependence of $\upsilon$ versus $\sqrt{\rho}$, as
observed in the numerical experiments. Moreover, 
the coefficient $\gamma_t$ is expected to decay when increasing the cutoff in
affinity $\alpha_c$, in agreement with the numerics.  
For $\rho >> 1$,  Eq.~\eqref{standard} implies:
\begin{equation}
 \upsilon = \sqrt{\frac{1}{12}}\, ,
 \label{large_rho}
\end{equation}
thus returning the correct result. 

To derive Eq.~\eqref{standard} let us suppose that at time  $t$ 
the death/birth process takes place and the system
experience an injection of new  
individuals. Label with $\upsilon_t$ the standard deviation of the agents
opinion distribution   
$f^t(O)$, at time $t$. It is reasonable to assume that $f^t(O)$  is centered
around $1/2$.  
After $T$ interactions between agents, when the next
perturbation will occur ($M$ agents are randomly removed from the
  community and replaced by $M$ new actors with random opinion and affinity scores) the  
distribution has been already modified, because of the underlying dynamical
mechanism specified through 
Eqs.~\eqref{opinion}. More concretely, the opinions slightly 
converge around the peak value $1/2$, an effect that certainly translates into a 
reduction of the associated standard deviation. To provide a quantitative estimate of such phenomenon, we 
recall that in the relevant $(O,t)$ plan, the convergence process fills an ideal 
{\it triangular} pattern, whose height measures $T_c$. This topological observation enables us to put forward the following linear ansatz:
 \begin{eqnarray}
\upsilon^{conv}_{t+T} = \upsilon_{t}\left(1  - \frac{T}{T_c}\right) \, ,
  \label{standard_t_plus_T_a}
\end{eqnarray}
where $\upsilon^{conv}_{t+T}$ labels the standard deviation just before the insertion of the next pool of incoming agents
\footnote{Numerical simulations (not reported here) show that in the closed model, the
  standard deviation of the opinions' distribution exhibits a exponential decay as a function of a power of time. 
  This latter is approximately interpolated by the proposed linear
  relation~\eqref{standard_t_plus_T_a}, a choice which eventually allows us to 
  carry out the  analytical calculation resulting in expression~\eqref{standard} (see also
  the discussion in the appendix~\ref{sec:app}). Formally, 
  eq. (\ref{standard_t_plus_T_a}) applies only for $t=0$, when agents are
  populating the interval $[0,1]$ with a 
uniform distribution. During the subsequent evolution, the convergence 
still gives rise to a macroscopic triangular pattern but, now, the associated triangle height $T_c$ gets slightly reduced. At time $t$ agents 
are still confined in the relevant interval $[0,1]$ and experience a  certain degree of spreading, effect of the perturbation 
externally imposed. However, and especially for intermediate values of $\rho$, the progressive bunching opposes the birth/death disturb (which
would tend to restore the $t=0$ variance) and drives an instantaneous reduction of $T_c$, as $t$ grows. In the following, and to account for
this self-consistent effect not captured by analytical framework,  $T_c$ is hence regarded
as an effective parameter to be numerically adjusted: as commented below however, the best fit values of $T_c$ correlate well with direct measurements of
the convergence (aggregation) time of the unperturbed system, a finding which a posteriori confirms the plausibility of eq. 
(\ref{standard_t_plus_T_a}).}

Recalling that the newly inserted elements are uniformly distributed, and labeling with 
$G(\cdot ,\cdot)$ the opinion distribution (the two entries referring respectively to mean and the standard deviation), the updated 
variance is:
 \begin{eqnarray}
 \upsilon_{t+T}^2  &=& \frac{M}{N} \int^{1}_{0} \left(O-\frac{1}{2} \right)^2
 d O \notag \\&+& \frac{N-M}{N} \int^{1}_{0} G \left (\frac{1}{2},\upsilon^{conv}_{t+T} \right)
 \left(O-\frac{1}{2}\right)^2 dO \notag\\
 &=& \frac{M}{12N} + \frac{N-M}{N} \left(1-\frac{T}{T_c}\right)^2
 \upsilon_{t}^2\, .\nonumber
  \label{standard_reborn}
\end{eqnarray}

The asymptotic stationary solution correspond to\\
$\upsilon_{t+T}=\upsilon_{t}$, a 
condition that immediately leads to Eq.~\eqref{standard} when plugged
into~\eqref{standard_reborn}. The above analysis also suggests that the final
fate of the system is not affected by the time when the  
perturbation is first applied, $t_{in}$. This conclusion is also confirmed by
direct numerical inspection:   
The asymptotic value of the standard deviation $\upsilon$ does not depend on
$t_{in}$, but solely on $\rho$. Even in the extreme condition, when the 
death/birth perturbation is switched on after
the agents have 
already collapsed to the mean opinion $0.5$,  
one observes that, after a transient, the cluster spreads and the measured
value of  $\upsilon$ agrees with the theoretical  
prediction~\eqref{standard}.

\begin{figure}[htbp]
\centering
\includegraphics[width=9cm]{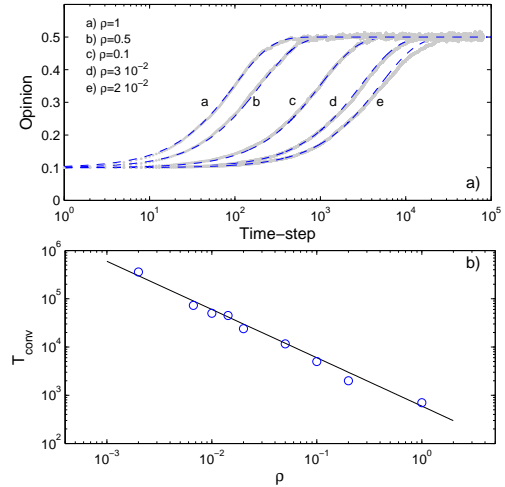}
\caption{a) The opinion as a function of time for the case where all
  the agents are initialized with the same opinion $O_o=0.1$. The solid (red)
  line represent  
the numerical data while the segmented (blue) line represents the theoretical prediction from Eq.~\eqref{mean_rOn2}. b) The transition time $T_{conv}$ as a 
function of the death density $\rho$. Symbols stand for the results of the simulations and the solid line represents the prediction~\eqref{T_{conv}}. Here $N=100$, $\alpha_c = \Delta O_c = 0 .5$, $\epsilon=0.001$ and $\sigma=0.07$.}
\label{fig4}
\end{figure}

Aiming at further characterizing the system dynamics, we also studied the case
where, initially, agents share the same belief $O_0$.  
The initial distribution of opinions is therefore a Dirac delta $f^0(O)=\delta
(O-O_o)$. Such condition is   
a stationary solution, for any given $O_o$ when the
death/birth process is inactivated.   
Conversely, when the death/birth applies, the system evolves toward a state, 
characterized by a single cluster (localized, if $\rho$ is small, uniformly spread over the allowed region as $\rho > 1$, see
preceding discussion),  centered in $O=0.5$ and with
standard deviation given by Eq.~\ref{standard}. It is also observed that the time needed by the system to complete the transition 
$T_{conv}$ depends on the value of $\rho$ and the critical affinity $\alpha_c$, see
Fig.~\ref{fig4}. A simple theoretical argument enables us to quantitatively explain these findings. The initial distribution of opinions is modified 
after the first death/birth event as:
 \begin{eqnarray}
 f^1(O) =\frac{M}{N}+\frac{N-M}{N} \delta (O-O_o)\, .
  \label{f_{1}}
\end{eqnarray}
The first term refers to the freshly injected actors,  while the second stands for the remaining Delta-distributed individuals. Hence, 
the mean opinion value reads:
\begin{equation}
 \bar{O}_1 =\int^{1}_{0}{f^1(O) O dO}=\frac{M}{2N}+\frac{N-M}{N} O_o\, .
  \label{mean_O1}
\end{equation}

We can suppose that between the occurrence of two consecutive perturbations (separated by $T$ iterations),
the group average opinion  
does not significantly change. Notice that the probability of interaction of a newborn agent with another belonging to the main group is in fact proportional to M/N. Moreover several consecutive encounters of this type are necessary to induce a macroscopic change of the averaged opinion.
Under this hypothesis the next death/birth event makes the
average opinion change as:  
\begin{equation}
 \bar{O}_{2} =\frac{M}{2N}+\frac{N-M}{N} \bar{O}_1\, .
  \label{mean_O2}
\end{equation}

After $n$ death/birth iterations, the opinion
mean value reads : 
\begin{equation}
 \bar{O}_{n} =\frac{M}{2N}+\frac{N-M}{N} \bar{O}_{n-1}\, .
  \label{mean_On}
\end{equation}
From Eq.~\eqref{mean_On} one easily gets that the
asymptotic equilibrium is reached for $\bar{O}_{\infty}=0.5$, as 
 found in our 
numerical experiments; in fact the following relation is
straightforwardly obtained: 
\begin{equation}
 \bar{O}_{n} =\left(\frac{N-M}{N}\right)^{n} O_o +
 \sum^{n-1}_{l=0}{\frac{M}{2N}\left(\frac{N-M}{N}\right)^l}\, ,
  \label{mean_rOn}
\end{equation}
being $O_o$ the initial common believe. By setting
$\alpha=\frac{N-M}{N}$ 
and $\beta=\frac{M}{2N}$, the solution of Eq.~\eqref{mean_rOn} reads: 
\begin{equation}
 \bar{O}_{n} =\beta \frac{1-\alpha ^{n}}{1-\alpha}+ \alpha ^{n} O_o\, ,
  \label{mean_rOn2}
\end{equation}
whose asymptotic solution is given by $\bar{O}_{n}\rightarrow
  \bar{O}_{\infty}=0.5$. 

Expression~\eqref{mean_rOn2} reproduces quite well the dynamics of the cluster
mean, as seen in the simulations. The adequacy of~\eqref{mean_rOn2} is in fact
clearly demonstrated in Fig.~\ref{fig4}a. Let us define the convergence time
$T_{conv}$  as the number of iterations needed to bring the average opinion
$\epsilon$ close to its asymptotic value $1/2$. Solving  Eq.~\eqref{mean_rOn2}
for $n$ and recalling that  $T_{conv}=nT$ yield: 
\begin{equation}
T_{conv} =  T  \log_{\alpha} \left[\frac{1}{|O_o - \frac{1}{2}|}(\epsilon - \frac{\beta}{1-\alpha }) \right]\, ,
  \label{T_{conv}}
\end{equation}
The above estimate is in excellent agreement with the numerical results reported in Fig.~\ref{fig4}b.  
\section{Conclusions}
\label{sect:conclusion}
In this paper we have discussed the process of opinion making in an
open group of interacting subjects. The model postulates the coupled
dynamical evolution of both individuals' opinion and mutual affinity,
according to the rules formulated in~\cite{Bagnoli_prl}. At variance
with
respect to the toy---model~\cite{Bagnoli_prl}, the system is now open
to contact with an external reservoir of potentially interacting
candidates. Every $T$ iterations the $M$ agents are instantaneously
replaced by newborn actors, whose opinion and affinity scores
are randomly generated according to a pre--assigned (here uniform)
probability distribution. The ratio $\rho=M/T$, here termed departure
density
plays the role of a control parameter. The occurrence of a 
transition is found which separates between two macroscopically
different
regimes: For large values of the so--called social temperature the
system collapses to a single cluster in opinion, while in the opposite
regime a fragmented phase is detected. The role of $\rho$ is
elucidated and shown to enter in the critical threshold as a linear
contribution.
Two phenomena are then addressed, with reference to the single
clustered phase. On the one side, the external perturbation, here
being hypothesized to mimic a death/birth process, induces
a spreading 
of the final cluster. The associated variance is numerically shown to
depend on the density amount
$\rho$, the functional dependence being also analytically explained.
On the other side, we also show that the birth/death events
imposed at a 
constant pace can produce the progressive migration of a cluster,
initially localized around a given opinion value. A theoretical
argument is also developed to clarify this finding. As a general
comment, we should emphasize that the effect of opening up the system to
external influences changes dramatically its intrinsic dynamics
revealing peculiar, potentially interesting, features which deserves to
be further explored.

\appendix
\section{Appendix}
\label{sec:app}

This appendix is devoted to discussing a straightforward extension of the above analysis to the
case of the original  Deffuant et al. model, which is made open via a birth/death 
mechanism as outlined above. The interested reader can consult~\cite{Deffuant} for a detail account 
onto the closed model specifications. We shall here solely recall that the Deffuant's setting 
$\mu=1/2$ and $d=1$, is formally recovered by setting 
$\alpha_c=0$ into the affinity model.  

In the closed Deffuant's setting, assuming $d=1$, the standard deviation of the opinion distribution
decays as an exponential function~\cite{bnks}, namely: 
\begin{equation}
  \label{eq:stddeff}
  v(t)=e^{-t/\tau_c}v(0)\, ,
\end{equation}
where $\tau_c$ plays the role of a {\it characteristic time}. Dedicated
numerical simulations, relative to the case study $d=1$, return the
$\tau_c=191.52$.

\begin{figure}[htbp]
\centering
\includegraphics[width=9cm]{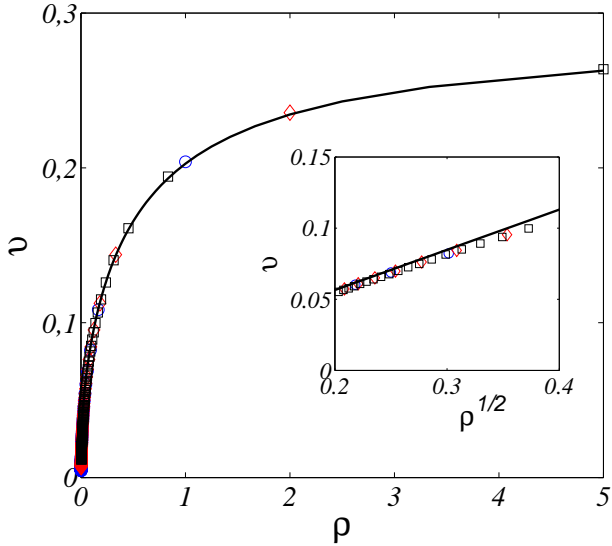}
\caption{Final standard deviation, $\upsilon$, of the opinion cluster 
  as a function of the departure density $\rho$. Each point results from 
  averaging out $100$ independent runs, each relative to $N=100$ agents. Symbols refer to
  numerical simulations performed with $d=1$, $\mu=1/2$ and different $M$ :
  ($\circ$) M=1, 
  ($\Diamond$) $M=2$ and ($\Box$) $M=5$. The solids line refers to the
  theoretical prediction~\eqref{standard_reborndeffstat} with $\tau_c=191.52$, this latter being  independently 
  estimated as discussed in the main text.} 
\label{fig:deff}
\end{figure}

Assume now that every $T$--steps the system opens: $M$ agents are randomly
removed. New actors enter the systems, their opinions being randomly
 sampled from a uniform distribution in the interval $[0,1]$. Let us denote $\rho=M/T$, the departure
density. Furthermore, label with  $\upsilon_{t+T}^{conv}$ the  
standard deviation of the opinions just before the insertion of the next pool of
incoming agents. Hence, in analogy with the preceding discussion, one can straightforwardly write the recursive relation: 

 \begin{equation}
 \upsilon_{t+T}^2  = \frac{M}{12N} + \left(1-\frac{M}{N}\right) e^{-2T/\tau_c}
 \upsilon_{t}^2\, ,
  \label{standard_reborndeff}
\end{equation}
whose asymptotic stationary solution corresponds to
 \begin{equation}
 \upsilon^2  = \frac{M}{12N}\frac{1}{1-\left(1-M/N\right)e^{-2T/\tau_c}}\, ,
  \label{standard_reborndeffstat}
\end{equation}
if $T$ is small enough. This result can be compared to the result of direct numerical simulation, returning an 
excellent agreement, as displayed in Fig.~\ref{fig:deff}. 

\begin{remark}
As a final remark,  let us observe that in the general case,
i.e. $\alpha_c~>~0$,  
an equation formally analogous to~\eqref{standard_reborndeff} can be derived,
by invoking the correct exponential  
ansatz (see main text). To obtain a closed analytical form for the asymptotic
stationary standard deviation 
we however decided to resort to a linear approximation for the opinions
convergence, as commented above.  
\end{remark}

Clearly, when starting from a preformed cluster of opinions the injection of
new actors determines an effective migration of the mean, also ruled by
Eq.~\eqref{mean_On} in the original Deffuant et al. scheme.


\begin{thebibliography}{99}
\bibitem{tempoconv} To compare the
    fitted convergence time $T_c$ with a direct measure of the aggregation
    time in  
    absence of birth/death perturbation, $\rho = 0$, we assumed the following
    stopping 
    criterion. We monitored the standard deviation $\upsilon_{sim} (t)$, of
    the simulated opinion distribution   
    and we measured the time when the latter gets smaller than a treshold
    amount $\epsilon_{thr}$ here set to  
    $0.001$ (i.e. $10$ times smaller than the initial inter-particle
    separation.).  The aggregation times thus obtained averaging over $50$
    indipendent runs, are $T_{agg}\sim 7100$ for $\alpha_c = 0.5$,
    $T_{agg}\sim 3000$ for $\alpha_c = 0.3$, and $T_{agg}\sim 1730$ for
    $\alpha_c = 0.0$, which, excepting the case $\alpha_c=0.3$, agree with the
    values of the fitted parameters $T_c$.
\end{thebibliography}
\end{document}